\documentclass[journal]{IEEEtran}
\usepackage{amsmath,amssymb,amsfonts,epsfig}
\usepackage{verbatim}
\usepackage{graphicx}
\usepackage{graphics}

\newcommand{\ber}{\begin{eqnarray}}
\newcommand{\eer}{\end{eqnarray}}

\newtheorem{theorem}{\noindent Theorem}
\newtheorem{lemma}{\noindent Lemma}

\newcommand{\be}{\begin{equation}}
\newcommand{\ee}{\end{equation}}

\newcommand{\eps}{\epsilon}

\newcommand{\nin}{\noindent}

\begin{document}
%
% paper title
\title{From finite-system entropy to entropy rate for a Hidden Markov Process}
%
%
% author names and IEEE memberships
% note positions of commas and nonbreaking spaces ( ~ ) LaTeX will not break
% a structure at a ~ so this keeps an author's name from being broken across
% two lines.
% use \thanks{} to gain access to the first footnote area
% a separate \thanks must be used for each paragraph as LaTeX2e's \thanks
% was not built to handle multiple paragraphs
\author{Or~Zuk,~\IEEEmembership{}  %{Member,~IEEE,}
        Eytan~Domany,~\IEEEmembership{} %{Member,~OSA,}
        Ido~Kanter~\IEEEmembership{} %{Member,~IEEE,}
        and~Michael~Aizenman~\IEEEmembership{} %{Member,~IEEE,} % <-this % stops a space
\thanks{E.D. and O.Z. are in the Department of Physics of Complex Systems,
Weizmann Inst. of Science.}
\thanks{I.K. is in the Department of Physics, Bar-Ilan Univ.}
\thanks{M.A. is in the Departments of Physics and Mathematics,
Princeton Univ.}}
% note the % following the last \IEEEmembership and also the first \thanks -
% these prevent an unwanted space from occurring between the last author name
% and the end of the author line. i.e., if you had this:
%
% \author{....lastname \thanks{...} \thanks{...} }
%                     ^------------^------------^----Do not want these spaces!
%
% a space would be appended to the last name and could cause every name on that
% line to be shifted left slightly. This is one of those "LaTeX things". For
% instance, "A\textbf{} \textbf{}B" will typeset as "A B" not "AB". If you want
% "AB" then you have to do: "A\textbf{}\textbf{}B"
% \thanks is no different in this regard, so shield the last } of each \thanks
% that ends a line with a % and do not let a space in before the next \thanks.
% Spaces after \IEEEmembership other than the last one are OK (and needed) as
% you are supposed to have spaces between the names. For what it is worth,
% this is a minor point as most people would not even notice if the said evil
% space somehow managed to creep in.
%
% The paper headers
\markboth{IEEE communication letters}{Shell
\MakeLowercase{\textit{et al.}}: Bare Demo of IEEEtran.cls for
Journals}
% The only time the second header will appear is for the odd numbered pages
% after the title page when using the twoside option.
%
% *** Note that you probably will NOT want to include the author's name in ***
% *** the headers of peer review papers.                                   ***

% If you want to put a publisher's ID mark on the page
% (can leave text blank if you just want to see how the
% text height on the first page will be reduced by IEEE)
%\pubid{0000--0000/00\$00.00~\copyright~2002 IEEE}

% use only for invited papers
%\specialpapernotice{(Invited Paper)}

% make the title area
\maketitle

\begin{abstract}
A recent result presented the expansion for the entropy rate of a
Hidden Markov Process (HMP) as a power series in the noise
variable $\eps$. The  coefficients of the expansion around the
noiseless ($\eps = 0$) limit were calculated up to 11th order,
using a conjecture that relates the entropy rate of a HMP to the
entropy of a process of finite length (which is calculated
analytically). In this communication we generalize and prove the
validity of the conjecture, and discuss the theoretical and
practical consequences of our new theorem.
\end{abstract}

% Note that keywords are not normally used for peerreview papers.

% For peer review papers, you can put extra information on the cover
% page as needed:
% \begin{center} \bfseries EDICS Category: 3-BBND \end{center}
%
% For peerreview papers, inserts a page break and creates the second title.
% Will be ignored for other modes.
\IEEEpeerreviewmaketitle

\section{Introduction}
% The very first letter is a 2 line initial drop letter followed
% by the rest of the first word in caps.
%
% form to use if the first word consists of a single letter:
% \PARstart{A}{demo} file is ....
%
% form to use if you need the single drop letter followed by
% normal text (unknown if ever used by IEEE):
% \PARstart{A}{}demo file is ....
%
% Some journals put the first two words in caps:
% \PARstart{T}{his demo} file is ....
%
% Here we have the typical use of a "T" for an initial drop letter
% and "HIS" in caps to complete the first word.
%%\PARstart{T}{his} demo file is intended to serve as a ``starter file"
%%for IEEE journal papers produced under \LaTeX\ using IEEEtran.cls version
%%1.6b and later.
% You must have at least 2 lines in the paragraph with the drop letter
% (should never be an issue)
%% May all your publication endeavors be successful.

\PARstart{L}{et} $\{X_N\}$ be a finite state stationary Markov
process over the alphabet $\Sigma = \{1,..,s\}$, and  let
$\{Y_N\}$ be its noisy observation (on the same alphabet). The
pair can be described by the Markov transition matrix $M = M_{s
\times s} = \{m_{ij}\}$ and  the emission matrix $R = R_{s\times
s}$, which yield the probabilities $P(X_{N+1} = j | X_N = i) =
m_{ij}$ and $P(Y_N = j | X_N = i) = r_{ij}$. We consider here the
case where the signal to noise ratio  (SNR) is small and  $M$ is
strictly positive ($m_{ij} > 0$)
  and thus has a unique stationary distribution.
  For the `high - SNR' regime one may  write $R = I + \eps T$,
  where $\eps > 0$ is some small number, $I$ is the identity matrix,
and the matrix $T = \{t_{ij}\}$ satisfies $t_{ii} < 0, \: t_{ij}
\geq 0 , \: \forall i \neq j$ and $\sum_{j=1}^{s}{t_{ij}} = 0$.

\nin The process $Y$ can be viewed as an observation of $X$
through a noisy channel. It is an example of a {\it Hidden Markov
Process (HMP)}, and is determined by the parameters $M$, $T$ and
$\eps$. More generally, {\it HMPs} have a rich and developed
theory, and enormous applications in various fields (see
\cite{Merhav,Rabiner}).

\nin An important property of $Y$  is its entropy rate. The
Shannon entropy rate of a stochastic process (\cite{Shannon})
measures the amount of 'uncertainty per-symbol'. More formally,
for $i \leq j$ let $[X]_i^j$ denote the vector $(X_i,..,X_j)$. The
entropy rate is defined as: \be \bar{H}(Y) = \lim_{N \to \infty}
\frac{H([Y]_1^N)}{N} \label{entropy_rate_def}\ee

\nin Where $H(X) = -\sum_X P(X) \log P(X)$; We will sometimes omit
the realization $x$ of the variable $X$, so $P(X)$ should be
understood as $P(X = x)$. For a stationary process the limit in
(\ref{entropy_rate_def}) exists and $\bar{H}$ can also be computed
via the conditional entropy  (\cite{Cover}) as: $\bar{H}(Y) =
\lim_{N \to \infty} H(Y_N | [Y]_1^{N-1})$.  Here $H(U|V)$
represents the conditional entropy, which for random variables $U$
and $V$ is the average uncertainty of the conditional distribution
of $U$ conditioned on $V$, that is $H(U|V) = \sum_{v} P(U = u) H(U
| V=v) $. By the chain rule for entropy, it can also be viewed as
a difference of entropies, $H(U|V) = H(U,V) - H(V)$.  This
relation will be used below.

\nin There is at present no explicit expression for the entropy
rate of a {\it HMP} (\cite{Merhav,Jacquet}). Few recent works
(\cite{Jacquet, Weissman:02,ZKD1}) have dealt with finding the
asymptotic behavior of $\bar{H}$ in several regimes, albeit giving
rigorously only bounds  or at most second (\cite{ZKD1}) order
behavior. Here we generalize and prove a relationship, that was
posed in \cite{ZKD1} as a conjecture, thereby turning the
computation presented there, of $\bar{H}$ as a series expansion up
to 11th order in $\eps$, into a rigorous statement.

\section{Theorem Statement and Proof}
Our main result is the following:
\begin{theorem}
Let $H_N \equiv H_N(M,T,\eps) = H([Y]_1^N)$  be the entropy of a
system of length $N$, and let $C_N = H_N - H_{N-1}$.
Assume\footnote{\nin It is easy to show that the functions $C_N$
are differentiable to all orders in $\eps$, at $\epsilon =0$. The
assumption which is not proven here is that they are in fact
analytic with a radius of analyticity which is uniform in $N$, and
are uniformly bounded within some common neighborhood of $\epsilon
=0$} there is some (complex) neighborhood $B_{\rho}(0) \subset
\mathbb{C}$ of zero, in which the (one-variable) functions $\{ C_N
\}$ and $\bar{H}$ are analytic in $\eps$, with a Taylor expansion
given by: \be C_N(M,T,\eps) = \sum_{k=0}^{\infty} C_N^{(k)}
\eps^k, \quad \bar{H}(M,T,\eps) = \sum_{k=0}^{\infty} C^{(k)}
\eps^k \ee

\nin (The coefficients $C_N^{(k)}$ are functions of the parameters
$M$ and $T$. From now on we omit this dependence). Then: \be N
\geq \lceil \frac{k+3}{2} \rceil \Rightarrow C_N^{(k)} = C^{(k)}
\ee

\label{main_thm}
\end{theorem}

\nin $C_N$ is an upperbound (\cite{Cover}) for $\bar{H}$. The
behavior stated in Thm. \ref{main_thm} was discovered using
symbolic computations, but proven only for $k \leq 2$ , in the
binary symmetric case (\cite{ZKD1}). Although technically involved
, our proof is based on two simple ideas. First, we distinguish
between the noise parameters at different sites. We consider a
more general process $\{Z_N\}$, where $Z_i$'s emission matrix is
$R_i = I + \eps_iT$. The process $\{Z_N\}$ is determined by
$M$,$T$ and $[\eps]_{1}^N$. We define the following functions: \be
F_N(M,T,[\eps]_1^{N}) = H([Z]_1^N) - H([Z]_1^{N-1}) \ee

\nin Setting all the $\eps_i$'s equal reduces us back to the $Y$
process, so in particular $F_N(M,T, (\eps,..,\eps)) = C_N(\eps)$.

\nin Second, we observe that if a particular $\eps_i$ is set to
zero, the observation $Z_i$ equals the state $X_i$. Thus,
conditioning back to the past is 'blocked'. This can be used to
prove:

\begin{lemma}

Assume $\eps_j=0$ for some $1 < j < N$. Then:
$$
F_N([\eps]_1^N) = F_{N-j+1}([\eps]_{j+1}^N)
$$

\begin{proof}
\nin $F$ can be written as the sum: \be F_N = -\sum_{[Z]_1^N}
P([Z]_1^{N-1}) P(Z_N | [Z]_1^{N-1}) \log P(Z_N | [Z]_1^{N-1})
\label{F_cond_eq}\ee

\nin Here the dependence on $[\eps]_1^N$ and $M,T$ is hidden in the
probabilities $P(..)$. Since $\eps_j=0$, we must have $X_j = Z_j$,
and therefore (since $X$ is a Markov chain), conditioning further to
the past is 'blocked', that is: \be \eps_j = 0 \Rightarrow P(Z_N |
[Z]_1^{N-1}) = P(Z_N | [Z]_j^{N-1}) \label{blocking_cond} \ee

%\nin (Note that eq. \ref{blocking_cond} is true for $j < N$, and not
%for $j=N$).
\nin Substituting in eq. \ref{F_cond_eq} gives:
$$
F_N = -\sum_{[Z]_1^N} P([Z]_1^{N-1}) P(Z_N | [Z]_j^{N-1}) \log P(Z_N
| [Z]_j^{N-1}) =
$$
%$$
%-\sum_{Z_{j}^N} \left[ P(Z_N | [Z]_j^{N-1}) \log P(Z_N |
%[Z]_j^{N-1}) \sum_{[Z]_1^{j-1}} P([Z]_1^{N-1})  \right] =
%$$
\be -\sum_{[Z]_{j}^N} P([Z]_j^{N}) \log P(Z_N | [Z]_j^{N-1}) =
F_{N-j+1} \ee

\end{proof}

\label{F_cond_lemma}
\end{lemma}

\nin Let $\vec{k} = [k]_1^N$ be a vector with $k_i \in \{\mathbb{N}
\cup 0\}$. Define its 'weight' as $\omega(\vec{k}) = \sum_{i=1}^N
k_i$. Define also: \be F_N^{\vec{k}} \equiv \left.
\frac{\partial^{\omega(\vec{k})} F_N}{\partial
\eps_1^{k_1},..,\partial \eps_N^{k_N}}
 \right|_{\vec{\eps} = 0}
\ee

\nin As we now show, adding zeros to $\vec{k}$ leaves $F_N^{\vec{k}}$ unchanged : % the left of
\begin{lemma}
Let $\vec{k} = [k]_1^N$ with $k_1 \leq 1$. Denote $\vec{k}^{(r)}$ % of $\vec{k}$ and $r$ zeros
the concatenation: $\vec{k}^{(r)} =
(\underbrace{0,..,0}_r,k_1,..,k_N)$. Then:
$$
F_N^{\vec{k}} = F_{r+N}^{\vec{k}^{(r)}} \quad,  \forall r \in
\mathbb{N}
$$

\begin{proof}
\nin Assume first $k_1 = 0$. Using lemma \ref{F_cond_lemma}, we get:
$$
F_{N+r}^{\vec{k}^{(r)}}([\eps]_1^{N+r}) = \left.
\frac{\partial^{\omega(\vec{k}^{(r)})}
F_{r+N}([\eps]_1^{N+r})}{\partial \eps_{r+2}^{k_2},..,\partial
\eps_{r+N}^{k_N}}
 \right|_{\vec{\eps} = 0} =
$$
\be
 \left.
\frac{\partial^{\omega(\vec{k})} F_{N}([\eps]_{r+1}^{N+r})}{\partial
\eps_{r+2}^{k_2},..,\partial \eps_{r+N}^{k_N}}
 \right|_{\vec{\eps} = 0} = F_N^{\vec{k}}([\eps]_{r+1}^{r+N})
 \label{k_is_zero_eq}
\ee

\nin The case $k_1 = 1$ is reduced back to the case $k_1=0$ by
taking the derivative. We denote ${[Z]_1^N}^{(j \to r)}$ the vector
which is equal to $[Z]_1^N$ in all coordinates except on coordinate
$j$, where $Z_j = r$. Using eq. \ref{k_is_zero_eq}, we get:
$$
F_{N+1}^{\vec{k}^{(1)}}([\eps]_1^{N+1}) = \left.
\frac{\partial^{\omega(\vec{k})-1}}{\partial \eps_3^{k_2} \dots
\partial \eps_{N+1}^{k_N}} \left[ \left. \frac{\partial
F_{N+1}}{\partial \eps_2} \right|_{\eps_2 = 0} \right]
\right|_{\vec{\eps} = 0} =
$$
$$
\frac{\partial^{\omega(\vec{k})-1}}{\partial \eps_3^{k_2} \dots
\partial \eps_{N+1}^{k_N}} \Biggl\{ - \sum_{r=1}^{s} t_{X_i r}
\sum_{[Z]_1^{N+1}}
$$
$$
  \left[ P({[Z]_1^{N+1}}^{(2 \to r)})
\log P(Z_{N+1} | [Z]_1^{N}) - \right.
$$
$$
\left. \left.  \left. P(Z_{N+1} | [Z]_1^{N}) P({[Z]_1^{N}}^{(2 \to
r)}) \right]  \right|_{\eps_2 = 0} \Biggr\} \right|_{[\eps]_1^{N+1}
= 0} =
$$
$$
\frac{\partial^{\omega(\vec{k})-1}}{\partial \eps_2^{k_2} \dots
\partial \eps_{N}^{k_N}} \Biggl\{ - \sum_{r=1}^{s} t_{X_i r}
\sum_{[Z]_1^{N}}
$$
$$
  \left[ P({[Z]_1^{N}}^{(1 \to r)})
\log P(Z_{N} | [Z]_1^{N-1}) - \right.
$$
\be \left. \left.  \left. P(Z_{N} | [Z]_1^{N-1}) P({[Z]_1^{N}}^{(1
\to r)}) \right]  \right|_{\eps_1 = 0} \Biggr\}
\right|_{[\eps]_1^{N} = 0} = F_{N}^{\vec{k}}([\eps]_1^{N})\ee

\end{proof}

\label{zero_tail_lemma}
\end{lemma}

\nin $C_N^{(k)}$ is obtained by summing $F_N^{\vec{k}}$ on all
$\vec{k}$'s with weight $k$: \be C_N^{(k)} =
\sum_{\vec{k},\omega(\vec{k})=k} F_N^{\vec{k}} \ee

\nin The next lemma shows that one does not need to sum on all such
$\vec{k}$'s, as many of them give zero contribution:

\begin{lemma}
Let $\vec{k} = [k]_1^N$. If $\exists i < j < N$, with $k_i \geq 1,
k_j \leq 1$, then $F_N^{\vec{k}} = 0$.

\begin{proof}
\nin Assume first $k_j = 0$. Using lemma \ref{F_cond_lemma} we get

$$
F_N^{\vec{k}} \equiv \left. \frac{\partial^{\omega(\vec{k})}
F_N(\vec{\eps})}{\partial \eps_1^{k_1},..,\partial \eps_N^{k_N}}
 \right|_{\vec{\eps} = 0} = \left. \frac{\partial^{\omega(\vec{k})}
F_{N-j+1}([\eps]_j^N)}{\partial \eps_1^{k_1},..,\partial
\eps_N^{k_N}}
 \right|_{\vec{\eps} = 0} =
$$
\be \frac{\partial^{\omega(\vec{k})-1}}{\partial \eps_1^{k_1},..,
\partial \eps_i^{k_i-1},.., \partial \eps_N^{k_N}} \left[\left.
\frac{\partial F_{N-j+1}([\eps]_j^N)}{\partial \eps_i} \right]
\right|_{\vec{\eps} = 0} = 0 \ee

%%%%%
%%%%%%%%%%%%%%%%%%%%%%%%%%%%%%%%%%%%%%%%%%%%%%%%%%%%%%%%%

\nin Assume now $k_j = 1$. Write the probability of $Z$:
$$
 P([Z]_1^N)  = \sum_{[X]_1^N} P([X]_1^N) P([Z]_1^N |
[X]_1^N) =
$$
\be \sum_{[X]_1^N} P([X]_1^N) \prod_{i=1}^N (\delta_{X_i Z_i} +
\eps_i t_{X_i Z_i})\ee

\nin where $\delta$ is Kronecker's delta. Differentiate with
respect to $\eps_j$:

$$
\left. \frac{\partial P([Z]_1^N)}{\partial \eps_j} \right|_{\eps_j =
0} =
$$
$$
\left. \sum_{[X]_1^N} \left[ P([X]_1^N) t_{X_j Z_j} \prod_{i \neq j}
(\delta_{X_i Z_i} + \eps_i t_{X_i Z_i}) \right] \right|_{\eps_j = 0}
=
$$
\be \left. \left\{ \sum_{r=1}^{s} t_{X_i r} P({[Z]_1^N}^{(j \to
r)}) \right\} \right|_{\eps_j = 0} \ee

\nin Using Bayes' rule $P(Z_N | [Z]_1^{N-1}) =
\frac{P([Z]_1^N)}{P([Z]_1^{N-1})}$, we get:

$$
\left. \frac{\partial P(Z_N | [Z]_1^{N-1})}{\partial \eps_j}
\right|_{\eps_j = 0} =
$$
$$
\frac{1}{P([Z]_1^{N-1})} \sum_{r=1}^{s} t_{X_i r} \left[
P({[Z]_1^N}^{(j \to r)}) - \right.
$$
\be \left. \left. P(Z_N | [Z]_1^{N-1} ) P({[Z]_1^{N-1}}^{(j \to r)})
\right] \right|_{\eps_j = 0} \ee

\nin This gives:
$$
\left. \frac{\partial [P([Z]_1^N) \log P(Z_N |
[Z]_1^{N-1})]}{\partial \eps_j} \right|_{\eps_j = 0} =
$$
$$
\sum_{r=1}^{s} t_{X_i r} \left\{ P({[Z]_1^N}^{(j \to r)}) \log
P(Z_N | [Z]_1^{N-1}) + \right.
$$
\be \left. \left. P({[Z]_1^N}^{(j \to r)})  - P(Z_N | [Z]_1^{N-1})
P({[Z]_1^{N-1}}^{(j \to r)})\right\} \right|_{\eps_j = 0} \ee

\nin And therefore:
$$
\left. \frac{\partial F_N}{\partial \eps_j} \right|_{\eps_j = 0} =
$$
$$
 -\sum_{r=1}^{s} t_{X_i r} \Biggl\{ \sum_{[Z]_1^N} \left[
P({[Z]_1^N}^{(j \to r)}) \log P(Z_N | [Z]_1^{N-1}) - \right.
$$
$$
 \left. \left. P(Z_N | [Z]_1^{N-1}) P({[Z]_1^{N-1}}^{(j \to
r)}) \right] \Biggr\} \right|_{\eps_j = 0} =
$$
%$$
%\left. \frac{\partial F_N}{\partial \eps_j}  \right|_{\eps_j = 0} =
%$$
$$
\Biggl\{ -\sum_{r=1}^{s} t_{X_i r} \sum_{[Z]_j^N} \left[
P({[Z]_j^N}^{(1 \to r)}) \log P(Z_N | [Z]_j^{N-1}) - \right.
$$
\be \left.  \left. P(Z_N | [Z]_j^{N-1}) P({[Z]_j^{N-1}}^{(1 \to r)})
\right] \Biggr\} \right|_{\eps_1 = 0} \label{F_N_deriv_eq} \ee

\nin The latter equality comes from using  eq.
\ref{blocking_cond}, which 'blocks' the dependence backwards. Eq.
\ref{F_N_deriv_eq} shows that $\eps_i$  does not appear in $\left.
\frac{\partial F_N}{\partial \eps_j} \right|_{\eps_j = 0}$ for $i
< j$, therefore $\left. \frac{\partial^{k_i+1} F_N}{\partial
\eps_i^{k_i}
\partial \eps_j} \right|_{\eps_j = 0}  = 0$ and $F_N^{\vec{k}} = 0$.

\end{proof}
\label{No_hole_strong_lemma}
\end{lemma}

\nin We are now ready to prove our main theorem:

\begin{proof}

\nin Let  $\vec{k} = [k]_1^N$ with $\omega(\vec{k}) = k$. Define its
'length'  as $l(\vec{k}) = N+1-\min_{k_i > 1} \{i \}$. It easily % (from right, considering only entries larger than one)
follows from lemma \ref{No_hole_strong_lemma} that $F_N^{\vec{k}}
\neq 0 \Rightarrow l(\vec{k}) \leq \lceil \frac{k+3}{2} \rceil - 1$.
Thus, according to lemma \ref{zero_tail_lemma}:% we have

\be F_N^{\vec{k}} = F_{\lceil \frac{k+3}{2} \rceil}^{(k_{N-\lceil
\frac{k+3}{2} \rceil+1},..,{k_N})} \ee

\nin for all $\vec{k}$'s in the sum. Summing on all $F_N^{\vec{k}}$
with the same 'weight' gives $C_N^{(k)} = C_{\lceil \frac{k+3}{2}
\rceil}^{(k)}, \: \forall N
> \lceil \frac{k+3}{2} \rceil$. But from the analyticity of $C_N$ and $\bar{H}$
near $\eps=0$ it follows that $\lim_{N \to \infty} C_N^{(k)} =
C^{(k)}$, therefore $C_N^{(k)} =
C^{(k)}, \: \forall N \geq \lceil \frac{k+3}{2} \rceil $. \\

\end{proof}
\label{strong_thm}

\section{Conclusion}
Our main theorem sheds light on the connection between finite and
infinite chains, and gives a practical and straightforward way to
compute the entropy rate as a series expansion in $\eps$ up to an
arbitrary power. The surprising 'settling' of the expansion
coefficients $C_N^{(k)} = C^{(k)}$ for $N \geq \lceil
\frac{k+3}{2} \rceil$, hold for the entropy. For other functions
involving only conditional probabilities (e.g. relative entropy
between two {\it HMPs}) a weaker result holds: the coefficients
'settle' for $N \geq k$. One can expand the entropy rate in
several parameter regimes. As it turns out, exactly the same
'settling' as was proven in Thm. \ref{main_thm} happens in the
'almost memoryless' regime, where $M$ is close to a matrix which
makes the $X_i$'s i.i.d. This and other regimes, as well as the
analytic behavior of the {\it HMP} (\cite{Marcus}), will be
discussed elsewhere.

% if have a single appendix:
%\appendix[Proof of the Zonklar Equations]
% or
%\appendix  % for no appendix heading
% do not use \section anymore after \appendix, only \section*
% is possibly needed

% use appendices with more than one appendix
% then use \section to start each appendix
% you must declare a \section before using any
% \subsection or using \label (\appendices by itself
% starts a section numbered zero.)
%
% Use this command to get the appendices' numbers in "A", "B" instead of the
% default capitalized Roman numerals ("I", "II", etc.).
% However, the capital letter form may result in awkward subsection numbers
% (such as "A-A"). Capitalized Roman numerals are the default.
%\useRomanappendicesfalse
%

%\appendices
%\section{Proof of the First Zonklar Equation}
%Appendix one text goes here.

% you can choose not to have a title for an appendix
% if you want by leaving the argument blank

% use section* for acknowledgement
\section*{Acknowledgment}
% optional entry into table of contents (if used)
%\addcontentsline{toc}{section}{Acknowledgment}
M.A. is grateful for the hospitality shown him at the Weizmann
Institute, where his work was supported by the Einstein Center for
Theoretical Physics and the Minerva Center for Nonlinear Physics.
The work of I.K. at the Weizmann Institute was supported by the
Einstein Center for Theoretical Physics. E.D. and O.Z. were
partially supported by the Minerva Foundation and by the European
Community's Human Potential Programme under contract
HPRN-CT-2002-00319, STIPCO.

\end{document}